\documentclass[conference]{IEEEtran}
\usepackage{xcolor} %,caption
\usepackage{xspace}

\colorlet{shadecolor}{yellow}
\usepackage[pdftex]{graphicx}
\DeclareGraphicsExtensions{.pdf,.jpeg,.png}

\usepackage[cmex10]{amsmath}
\usepackage{array}
\usepackage{url}
\usepackage[english]{babel}
\usepackage{blindtext}
\usepackage{subfig}

\newcommand{\nnn}{\texttt{NNN}\xspace}
\newcommand{\nnna}{\texttt{NNN-5}\xspace}
\newcommand{\nnnb}{\texttt{NNN-7}\xspace}
\newcommand{\nn}{\texttt{NN}\xspace}
\newcommand{\nna}{\texttt{NN-5}\xspace}
\newcommand{\nnb}{\texttt{NN-7}\xspace}
\newcommand{\cg}{\texttt{CG}\xspace}
\newcommand{\cga}{\texttt{CG-5}\xspace}
\newcommand{\cgb}{\texttt{CG-7}\xspace}
\newcommand{\pcg}{\texttt{PCG}\xspace}
\newcommand{\pcga}{\texttt{PCG-5}\xspace}
\newcommand{\pcgb}{\texttt{PCG-7}\xspace}
\newcommand{\scg}{\texttt{SCG}\xspace}
\newcommand{\scga}{\texttt{SCG-5}\xspace}
\newcommand{\scgb}{\texttt{SCG-7}\xspace}
\newcommand{\shcg}{\texttt{ShCG}\xspace}
\newcommand{\shcga}{\texttt{ShCG-5}\xspace}
\newcommand{\shcgb}{\texttt{ShCG-7}\xspace}

\begin{document}

\title{Neural Acceleration of Incomplete Cholesky Preconditioners}
\author{\IEEEauthorblockN{Joshua Dennis Booth}
\IEEEauthorblockA{Computer Science \\
University of Alabama in Huntsville \\
Huntsville, Alabama 358999 \\
joshua.booth@uah.edu}
\and
\IEEEauthorblockN{Hongyang Sun}
\IEEEauthorblockA{Electrical Engineering \& Computer Science \\
University of Kansas \\
Lawrence, KS 66045}
\and
\IEEEauthorblockN{Trevor Garnett} 
\IEEEauthorblockA{Computer Science \\
University of Alabama in Hunstville \\
Huntsville, Alabama 35899
}
}

% ====================================================================
\maketitle

% === ABSTRACT ====================================================================
% =================================================================================
\begin{abstract}
The solution of a sparse system of linear equations is ubiquitous in scientific applications. 
Iterative methods, such as the Preconditioned Conjugate Gradient method (PCG), are normally chosen over direct methods due to memory and computational complexity constraints.
However, the efficiency of these methods depends on the preconditioner utilized.
The development of the preconditioner normally requires some insight into the sparse linear system and the desired trade-off of generating the preconditioner and the reduction in the number of iterations.
Incomplete factorization methods tend to be black box methods to generate these preconditioners but may fail for a number of reasons.
These reasons include numerical issues that require searching for adequate scaling, shifting, and fill-in while utilizing a difficult to parallelize algorithm.
With a move towards heterogeneous computing, many sparse applications find GPUs that are optimized for dense tensor applications like training neural networks being underutilized. 
In this work, we demonstrate that a simple artificial neural network trained either at compile time or in parallel to the running application on a GPU can provide an incomplete sparse Cholesky factorization that can be used as a preconditioner.
This generated preconditioner is as good or better in terms of reduction of iterations than the one found using multiple preconditioning techniques such as scaling and shifting.
Moreover, the generated method also works and never fails to produce a preconditioner that does not reduce the iteration count. 
\end{abstract}

% === KEYWORDS ====================================================================
% =================================================================================

% \fi
\IEEEpeerreviewmaketitle

% ====================================================================
% ====================================================================
% ====================================================================

% === I. INTRODUCTION =============================================================
% =================================================================================
\section{Introduction}
\label{sec:intro}
Scientific applications in many domains depend on the solution of sparse linear systems.
While traditional Gaussian elimination-based methods (i.e., direct methods for factorizing) offer the best numerical stability, iterative methods dominate implementations in distributed-memory systems and accelerators (i.e., GPUs).
The reason is that iterative methods require less memory and utilize vectorized operations in many cases.
A common iterative method for symmetric positive definite (SPD) systems is the Preconditioned Conjugate Gradient method (PCG)~\cite{cg,saadbook} as it only relies on sparse matrix-vector multiplication (SpMV) and sparse triangular solve (Stri).
However, the question of generating ``good" preconditioners for a generic SPD system can be more of an art than a science.
Incomplete sparse factorization methods, e.g., incomplete Cholesky, are black box methods that are typically used to generate these preconditioners.
These methods normally require trying techniques such as scaling, shifting, and identifying fill-in to achieve the desired reduction in iterations.
However, the algorithm of incomplete factorization tends to be difficult to parallelize due to the low computational intensity, i.e., the ratio of the number of floating point operations to memory accesses~\cite{javelin}. 
In this work, we explore the use of neural acceleration to generate a preconditioner in order to automate this process for scientific application users and better utilize the heterogeneous computing environments common in high-performance computing. 

Neural acceleration is the method of replacing key computationally expensive kernels in code with very simple and cheap artificial neural networks~\cite{gpnn,gpunn}.
The idea is aimed at modern workloads that execute on heterogeneous systems.
These systems commonly contain GPU accelerators that are optimized for the dense tensor computations utilized in training neural networks. 
These neural networks are small enough to easily be trained and executed during compile time or in parallel to the application. 
In order to utilize neural acceleration, the coder flags functions that are computationally expensive but may not suffer from being approximated by a neural network. 
Normally, a small amount of information is provided about the function, such as expected inputs, outputs, and computational flow.
A cheap neural network is then trained on the GPU.
At the time of execution, the neural network can be utilized on the GPU in place of the function call.

The use of neural acceleration for generating a preconditioner is ideal as these preconditioners are normally an approximation of the input sparse matrix and the computation of the preconditioner can be less than ideal due to low computational intensity. 
Moreover, many techniques are being developed for sparse computations of neural networks on GPUs due to the growing importance of graph neural networks~\cite{fastgnn}. 
However, the question exists if a neural network could be utilized in this relatively simple manner.
In order to explore this, we consider the case of generating a preconditioner for PCG based on the computational flow of incomplete Cholesky factorization.

In particular, we explore a neural acceleration method for generating an incomplete Cholesky factorization with zero fill-in that performs as good as or better than a tuned incomplete Cholesky factorization without the overhead of trying different techniques.
As such, our method works as a black box for a wide range of sparse matrices and works for our own test suite of sparse matrices while most traditional methods fail in some cases. 
Our contributions are as follows:
\begin{itemize}
    \item A method to generate a high-quality preconditioner with a given sparsity pattern using neural networks (Section~\ref{sec:alg});
    \item A comparison of our method to other standard incomplete factorization methods that utilize a given sparsity pattern (Section~\ref{sec:exp});
    \item An analysis of timing costs to justify the use of neural acceleration (Section~\ref{sec:cost});
    %\item An examination of future extension of this work (Section~\ref{sec:future}). 
\end{itemize}

\section{Background and Related Work}
\label{sec:background}
This section provides a background into sparse incomplete factorization used as a preconditioner and the concept of neural acceleration.

\subsection{Sparse Preconditioning}
%\noindent \textbf{Traditional Methods.}
\subsubsection{Traditional Methods}
Most traditional methods focus on providing a universal robust method to generate a preconditioner for iteration methods such as PCG~\cite{cg} and GMRES~\cite{gmres}.
The most common of these is incomplete decomposition as it fits a wide array of unstructured systems.
There are two forms of these incomplete methods, i.e., $IChol(k)$  and $IChol(\tau)$~\cite{duff,javelin,saadbook}.
The former, $IChol(k)$, is based on the level of fill-in, i.e., zero elements becoming nonzeros during factorization, of a sparse matrix.
Here it is common to utilize $IChol(0)$, i.e., allowing no fill-in and thus having the same nonzero pattern as the input matrix, or $IChol(1)$ as these have a small memory footprint.
The second method, $IChol(\tau)$, is based on the numerical value of elements related to the off-diagonal.
Off-diagonal elements that are smaller than some $\tau$ or $\tau|a_{ii}|$ are dropped.
Additionally, some combination of these two, i.e., $IChol(k,\tau)$, can be utilized. 
However, the nonzero values in all these methods are derived from the truncated factorization method.
This means that errors (e.g., $\epsilon$ that is removed by dropping a nonzero earlier in the incomplete factorization)  might have a large impact on some values later on in the computation. 

Due to loss of precision from dropping nonzeros, many times the incomplete factorization may fail even though the input matrix is SPD.
In these cases, several options exist.
The first is to simply allow for more fill-in, but this will suffer from increased computation and memory costs.
The other two methods try to deal with the numerical issues directly~\cite{shiftcg}.
The first numerical method is applying scaling to the sparse matrix.
Sparse matrices from multiphysics problems can have element values that come from a large distribution and result in diagonal values tending toward zero when updated.
Relating back to neural networks, scaling of data is very common because of the numerical values desired by optimization methods utilized in training.
The second numerical method is to shift the diagonal values by some small amount to prevent them from tending toward zero when updated.
The value of the shift should be large enough to prevent the incomplete factorization method from failing, but small enough that the preconditioner is close to the original sparse matrix.
However, both of these introduce two more parameters to consider while constructing a preconditioner. 

A last consideration exists in the form of a sparse matrix ordering.
The amount of fill-in during factorization is a factor or the nonzero pattern. 
Certain orderings, e.g., nested-dissection (ND)~\cite{nd} and reverse Cuthill-McKee (RCM)~\cite{rcm}, are known to reduce fill-in.
In theory, the reduction in fill-in should result in better preconditioners (i.e., a reduction in the number of iterations to converge) when a fixed number of nonzeros is applied (e.g., $ICHOL(0)$) as there should be less error due to truncation. 
While this idea holds in general, it does not hold for all orderings.
An example of this is approximate minimal degree (AMD)~\cite{amd} ordering which reduces fill-in but does not reduce iteration count to the same degree as RCM~\cite{duff,order1}.

%\noindent \textbf{Theoretical Methods.}
\subsubsection{Theoretical Methods}
The traditional incomplete factorization methods work well in practice, although they do require tuning parameters to match the desired convergence rate while trading off memory usage and time.
There have been attempts to build more theoretically constructed preconditioners.
The reason for this is both academic and due to concerns with performance.
One such method is the use of support graphs to construct preconditioners for M-matrices~\cite{supportgraph}.
In this method, sparse matrices are viewed with their graph representation (i.e., rows/columns as vertices and nonzero values as edges with weights).
The idea is to determine the importance of an edge or additional edges and what the weights of these edges should be using the metric of a matrix pencil or eigenvalue problem. 
While this works ideal for small input matrices, the idea tends not to scale to larger and more general sparse systems~\cite{support2}.
A more modern approach to this is preconditioners built from graph sparsification~\cite{grass}. 
This approach utilizes larger and more global information to remove edges and reweigh the graph representation to have a more ideal eigenvalue distribution.
However, both of these methods suffer from not working on a wide range of sparse matrices, and the algorithms are difficult to scale (i.e., graph and sparse eigenvalue computations). 

\begin{figure}[tbh]
    \centering
    \includegraphics[width=.45\textwidth]{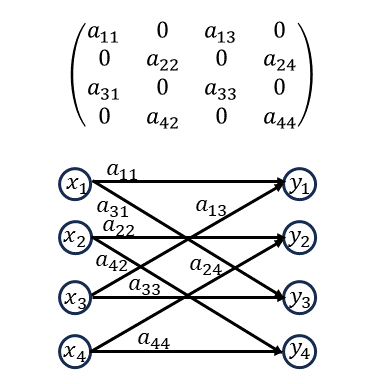}
    \caption{Neural network representation of $Ax=y$. The input nodes ($x_i$) represent the elements of vector $x$, the output nodes ($y_j$) represent the elements of vector $y$, and the edge weights are taken from nonzero elements of the sparse matrix $A$.}
    \label{fig:Ann}
\end{figure}

\subsection{Neural Networks and Acceleration}
The concept of utilizing neural networks to either solve or help solve systems of linear equations is not new. 
This is not surprising as there exists a direct relationship between solving a system and a general dense layer neural network.
Figure~\ref{fig:Ann} shows a visual representation for the matrix-vector multiplication ($Ax=y$). 
The nonzero values of the matrix are represented in the network as the edge weights while the input ($x_i$) and outputs ($y_j$) represent the input and output nodes.
No place is this connection seen more than in the foundational work related to online training such as Hopfield networks~\cite{Hopfield,wang, badHopfield, part1}.
These types of neural computations look at solving the system by finding the parameters of the connections (i.e., solving for $A^{-1}$) in an online manner (i.e., during runtime to include training).
In particular, they train the network by allowing for a fully connected network (i.e., all input nodes connected to all output nodes) and flip the inputs and outputs (i.e., $x_i \leftrightarrow y_j$).
Though there has been a lot of fundamental work in this area for things like embedded systems, these are designed for small dense networks with very specific restrictions to parameter distributions. 
One of the main reasons for this is the numerical instability of deriving an inverse.

In addition to these traditional approaches, modern work has started to analyze where neural networks can be used in solving sparse linear systems.
The work by Gotz and Anzt~\cite{hartwig} utilizes complex convolutional neural networks to identify blocking locations for generating block Jacobi preconditioners.  
The work demonstrates that a neural network can be used to identify good blocking for this type of preconditioner, but the neural network developed is very large with tens of thousands of parameters even for small problems (i.e., $dim(A) < 1000$)  and it would be difficult to fit into the framework of simple models utilized by neural acceleration. 
We also note that this type of problem, i.e., identifying blocking patterns, is very similar to cluster and edge detection done often by neural networks within the area of image processing.
In particular, the network utilized in this work is very close to that used in LeNet-5 for images. 

One modern neural acceleration approach to sparse linear systems considers the problem of identifying the sparse matrix ordering and calculation of fill-in~\cite{boothfillin}.
This work uses a simple graph neural network that represents the computational flow for the calculation of fill-in (i.e., column-by-column calculation).
They utilize neural acceleration to outperform traditional methods in CHOLMOD~\cite{cholmod,boothfillin} when utilizing a GPU. 
However, this work does not generate factorization or give insight into the problem of precondition generation for sparse iterative solvers. 
However, the fundamental takeaway from that work is that the search space for a neural network model used by neural acceleration should match the workflow.
This same principle of matching the workflow is utilized here.

\section{Neural Network Construction}
\label{sec:alg}

\subsection{Overview}
The current trend in neural networks is to construct a large (and most likely expensive) network that is very generic.
This means a large network would be trained (i.e., in a supervised manner) with a large training set of sparse matrices as inputs and ideal preconditioners as outputs.
While generic networks like these have many positive attributes, such as being able to be reused, they have a number of downfalls that make them less than ideal for neural acceleration and sparse linear algebra.
%We do note that there has been an attempt to use them for SpMV~\cite{}, but the results were subpar to more traditional methods~\cite{csrkgpu}.
The reasons are: (a) the training time normally outweighs traditional computational methods even when amortized over the number of uses, and (b) the training set for sparse linear algebra is very tiny.
In particular, there are very few different nonzero patterns and numerical values to construct a big enough training set to train a huge neural network model.
An example of this is demonstrated in the neural acceleration work related to graph ordering and fill-in~\cite{boothfillin}.

Moving away from these large generic networks, we outline how we construct our neural networks for incomplete Cholesky preconditioners based on the computational flow.
Many traditional neural network inspired approaches try to construct a preconditioner $M$ such that $M \approx A^{-1}$~\cite{part1,badHopfield,wang}.
In particular, the network itself becomes the output.
However, constructing an inverse directly is an error-prone task.
Numerical methods understand that $A^{-1}$ can be numerically unstable and will likely be dense.
This is one of the reasons sparse incomplete factorizations make sense.
Sparse factorizations, in general, are used to combat these problems by replacing $M$ with stable sparse triangular solves.
As this is the standard workflow, the neural network model should have the same workflow, i.e., the neural network should have the pattern of the incomplete factorization.
The problem can be further shrunk by fixing the nonzero pattern of the factorization, e.g., $ICHOL(k)$.

\begin{figure}[tbh]
    \centering
    \includegraphics[width=.45\textwidth]{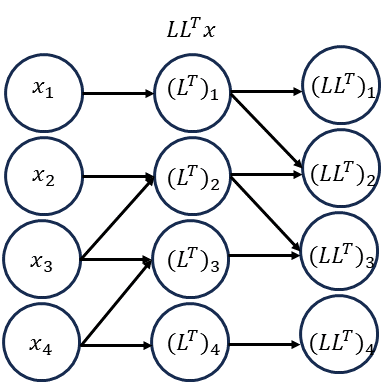}
    \caption{Neural network model of $LL^{T}x$. Here, the nonzero pattern (i.e., the edges) is based on the same nonzero pattern of Figure \ref{fig:Ann}. However, one hidden layer is added to the product of $L^{T}x$.  While the edges themselves are fixed based on the provided pattern, their numerical value will change based on training via backpropagation. }
    \label{fig:LLnn}
\end{figure}

Figure~\ref{fig:LLnn} visualizes a very simple sparse two-layer model. The edges in the first layer represent the nonzeros in $\hat{L}^T$, i.e., the transpose of a lower triangular matrix, and the edges of the second layer represent the nonzero in $\hat{L}$, i.e., they are ordered in the manner they would be applied to $\hat{L}\hat{L}^{T}x$.
Our decision to utilize a fixed ordering and nonzero pattern lends itself to us because we are able to know what edges we desire to use for our model.
In modern theoretical approaches to constructing incomplete sparse Cholesky (e.g., support-graph and sparsification),  both the edges and weights are flexible.
In more traditional level-based methods, the edges are fixed and the weights are calculated based on some truncation of the standard factorization method.
In this method, the edges are fixed but the weights for them will be calculated based on the model.
Note that this could be applied to any level of $ICHOL(k)$.
In our analysis of quality (Section~\ref{sec:exp}), we still restrict ourselves to $ICHOL(0)$.
We discuss this more in the next subsection.

\begin{figure*}[tbh]
    \centering
    \includegraphics[width=.95\textwidth]{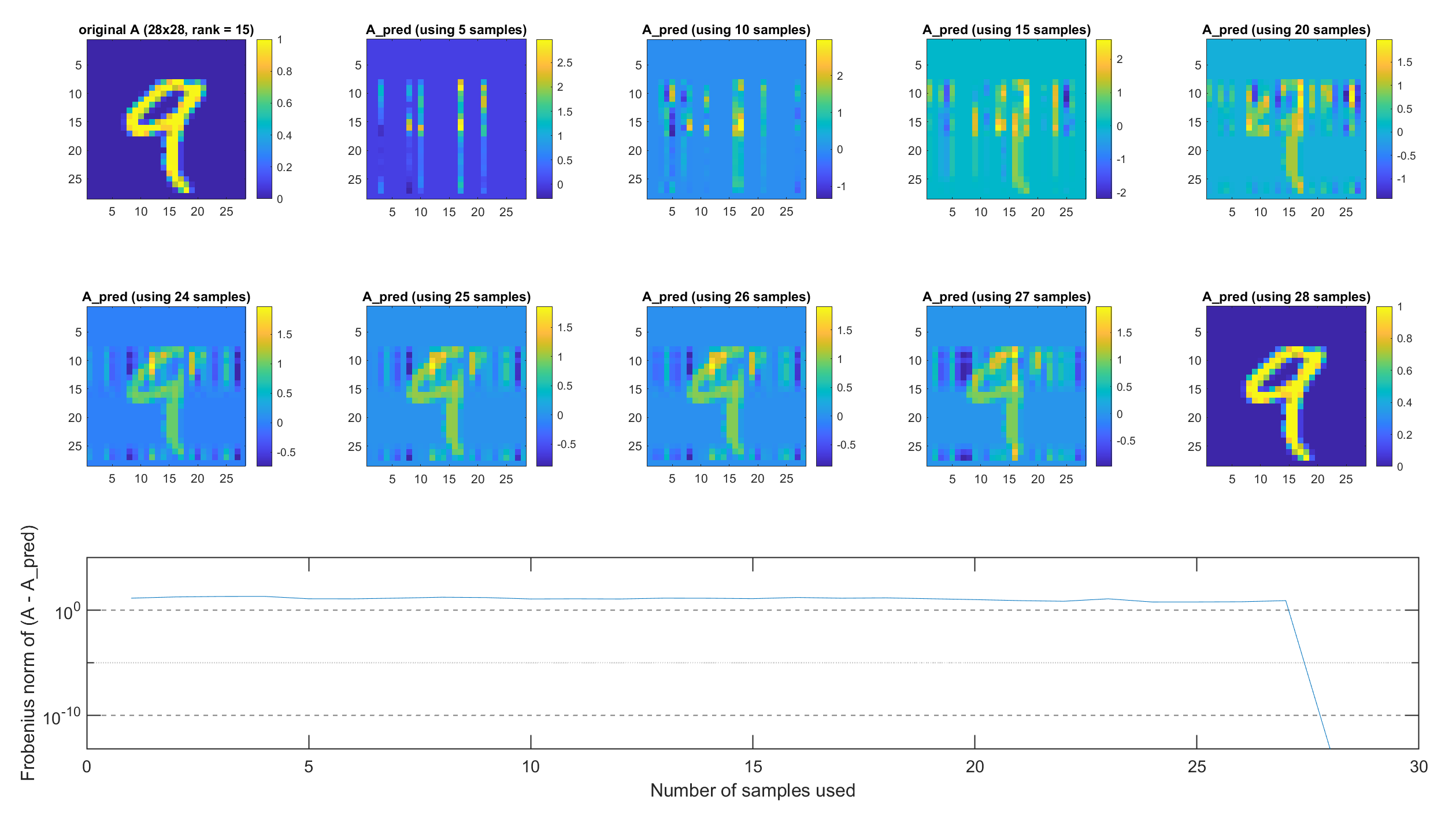}
    \caption{Reconstruction of an MNIST image (number nine) as a matrix with increasing number of samples. The first two rows of images provide the visual reconstructions and the bottom figure provides the error in terms of the Frobenius norm of the difference between the original and reconstructed images. We note that it is difficult to even make out the number at fewer than 24 samples and that the error norm only decreases at the point of 28 samples (i.e., the number of samples equals the dimension of the image)}
    \label{fig:matlab}
\end{figure*}

This simple model lends itself well to the current methods of training, namely backpropagation.
While methods like those in Hopfield networks~\cite{part1,badHopfield,wang} can be trained online, they require smaller networks with certain distributions.
In fact, they generally boil down to the gradient descent method for training.
With the current power of GPUs acting as accelerators, backpropagation methods that involve solving an optimization problem make sense. 
In particular, the objective function of such an optimization method could be written as:
\begin{equation}
\min ||Y-\hat{L}\hat{L}^T X||_2^{2}
\label{eq:mse}
\end{equation}
Here, we minimize the objective function by finding the numerical values for a fixed set of elements of $L$ using $Y$ and $X$ training values calculated via $Y=AX$, and the 2-norm represents the mean square error. 

Therefore, the neural acceleration method could take in the sparse matrix $A$ and generate samples $X$ and $Y$ in order to train $L$.
In our experimental results, we demonstrate that the number of samples needed is relatively small (i.e., $\sqrt{N}$ where $N=dim(A)$).
For the output, the method could either output $L$ to be used by the problem in its iterative solver package or function pointers to apply sparse triangular solve for this on the GPU where it was generated.

\subsection{Discussion}
\label{sec:discussion}
Several points of this method and implementation stand out in a manner that requires more discussion. 
For our method, we suggest using a predefined nonzero pattern and putting it more in the same grouping as $ICHOL(k)$.
This suggestion has several attributes.
First, a fixed pattern is normally inexpensive to compute and would have zero computational cost and limited memory overhead if $ICHOL(0)$ is utilized.
Second, a non-fixed pattern (e.g., $ICHOL(\tau)$) would require additional training parameters (i.e., meta parameters related to sparsification), and would require a much more expensive training search space.
We do not perceive this as a limitation of the neural acceleration technique as many fast parallel incomplete factorization methods make the same assumption~\cite{javelin,fast}, and the goal is to achieve a fast approximation.

The second point that deserves discussion is training cost. 
Section~\ref{sec:cost} provides an empirical analysis of this cost for our less-than-optimal training implementation.
The training cost would depend on both the numerical optimization method used and the number of iterations needed to construct a good approximation. 
The issue with this cost is that better and often more expensive numerical optimization methods require fewer iterations.
In our experimentation, we utilize stochastic gradient descent (SGD) and AdaGrad~\cite{adagrad} as our numerical optimizers.
SGD is cheaper per iteration.
We note that we would not even attempt training with SGD  in practice except to demonstrate the versatility of the model.
SGD utilizes the same update rate for each parameter.
AdaGrad utilizes the second-order information for updating to provide adaptive learning rates for each parameter.
AdaGrad is commonly used for training deep learning models with sparse gradients (e.g., recurrent neural networks and transformers). 
Overall, SGD normally requires about $N$ iterations to achieve the same quality as AdaGrad using $\sqrt{N}$ iterations. 
With AdaGrad, we find that the maximum time to train any of our test sparse neural networks is very small, and the goal of neural acceleration is to construct an approximation in a timely manner (i.e., having a small search space) that requires the least input from the user.
We continue the decision of this along with choices in Section~\ref{sec:cost}.
Moreover, we only provide the results for AdaGrad in the results as the time to train using SGD for our test suite is too high. 

In terms of training time and number of samples, there is an example to consider.
We note that in the forward direction of $Ax=y$, where we are trying to reconstruct $A$ from samples $X$ and $Y$, we would need $N$ samples.
Figure~\ref{fig:matlab} demonstrates this with an image (of number nine) taken from the MNIST dataset \cite{mnist}.
This image can be viewed as a sparse matrix without full rank (i.e., the rank does not equal the dimension of $A$).
The samples from $X$ are taken from a random distribution to produce $Y$ values.
The reconstruction is done using the standard backsolve operation (i.e., with QR decomposition~\cite{golub,davisQR} when the size is less than $N$) in Matlab.
While this method does not provide the required incomplete factorization, it does give us a sense of how many samples we really should need to construct the incomplete factorization.
Even with a sparse matrix that is not full rank, the image remains difficult to see after 25 samples and the error is very high (i.e., $>1$).
Therefore, the number of samples should be $\leq N$ to be a successful method.

Lastly, we discuss the possibility of extending this method from $ICHOL$ to incomplete $LU$.
The largest issue that normally impacts the generalization of $ICHOL$ to incomplete $LU$ is the loss of stability that is normally handled with pivoting.
Some incomplete $LU$ will also generate a permutation matrix $P$ that provides the row permutations as a result of pivoting.
Allowing this type of incomplete $LU$ may be very expensive and difficult with neural networks, though other neural acceleration methods have considered permutations in general~\cite{boothfillin}.
Because pivot serializes factorization, most parallel packages try to avoid pivoting using some reordering methods (e.g., those that permute large entries to the diagonal)~\cite{javelin} or limit the search for a pivot to a smaller subblock~\cite{basker}.
The first choice would make our method more scalable in terms of performance and reduce the model search space.
We plan to demonstrate this in future work.

\section{Experimental Setup}
\label{sec:expsetup}

\subsection{Matrix Test Suite}
We test our method utilizing symmetric positive definite matrices taken from the SuiteSparse Matrix Collection~\cite{ufm}.
A list of the sparse matrices along with their key parameters is provided in Table~\ref{tab:matrix}.
These parameters include the matrix ID, the dimension of the sparse matrix (N), the number of nonzeros (NNZ), the average row density (Density) (i.e., NNZ/N), and if the matrix is symmetric diagonally dominate (SDD), i.e., $|a_{ii}| \geq \sum_{i\neq j} |a_{ij}|$.
The ID is used to identify the sparse matrix in the figures in later sections to save space and make them more readable.
A horizontal line is drawn between the matrices with IDs 12 and 13 to represent where the figure is broken into two.
We provide the column related to SDD to indicate what matrices even fall into the category where a more theoretical method could be utilized.
Support graph type theoretical problem require M-matrix which are a more restrictive group than SDD.
Some of the sparsification methods will work with SDD matrices.

Additionally, we consider both the natural ordering (i.e., the ordering provided by the input) and the RCM ordering in our quality analysis as this has been shown in the past to be a key factor in iteration count~\cite{duff}.
Therefore, it is interesting to examine if the ordering has an impact when the preconditioner is generated using our neural network model.

\begin{table}[h]
\centering
\caption{Matrix test suite. ID is the number used to identify the matrix below, N is the matrix dimension, NNZ is the number of nonzeros in the matrix, and density is the average number of nonzeros per row.}
    \label{tab:matrix}
    \begin{tabular}{|r| c | c | c | c | c|}
    \hline
    \textbf{ID} &\textbf{Matrix} & \textbf{N} &  \textbf{NNZ} & \textbf{Density} & \textbf{SDD}  \\
    \hline
    1 & minsurfo & $40,806$ & $206$k & 1.22$e$-4 & Y \\
    2 & cvxbqp1 & $50,000$ & $350$k & 1.40$e$-4 & N \\
    3 & gridgena & $48,962$ & $512$k & 2.15$e$-4 & N \\
    4 & cfd1 & $70,656$ & $1,826$k & 3.66$e$-4  & N \\
    5 & oilpan & $73,752$ & $2,149$k & 3.95$e$-4 & N \\
    6 & vanbody & $47,072$ & $2,329$k & 1.05$e$-3 & N \\
    7 & ct20stif & $52,329$ & $2,600$k & 9.50$e$-4 & N \\
    8 &nasasrb & $54,870$ & $2,677$k & 8.89$e$-4 & N \\
    9 & cfd2 & $123,440$ & $3,085$k & 2.03$e$-4 & N \\
    10 & s3dkt3m2 & $90,449$ & $3,686$k & 4.51$e$-4 & N \\
    11 & cant & $62,451$ & $4,007$k & 1.03$e$-3 & N \\
    12 & shipsec5 & $179,860$ & $4,599$k & 1.42$e$-4 & N\\ \hline
    13 & consph & $83,334$ & $6,010$k & 8.66$e$-4 & N \\
    14 & G3\_circuit & $1,585,478$ & $7,661$k & 3.05$e$-6 & Y \\
    15 & hood & $220,542$ & $9,895$k & 2.03$e$-4 & N \\
    16 & thermal2 & $1,228,045$ & $8,580$k & 5.69$e$-6 & N \\
    17 & af\_0\_k101 & $503,625$ & $17,551$k &  6.92$e$-5 & N \\
    18 & af\_shell3 & $504,855$ & $17,562$k & 6.89$e$-5 & N \\
    19 & msdoor & $415,863$ & $19,173$k & 1.11$e$-4 & N \\
    20 & StocF-1465 & $1,465,137$ & $21,005$k & 9.79$e$-6 & N \\
    21 & Fault\_639 & $638,802$ & $27,246$k & 6.68$e$-5 & N \\
    22 & inline\_1 & $503,712$ & $36,816$k & 1.45$e$-4 & N \\
    23 & PFlow\_742 & $742,793$ & $37,138$k & 6.73$e$-5 & N \\
    24 & ldoor & $952,203$ & $42,494$k & 4.69$e$-5 & N \\
    \hline
\end{tabular}
\end{table}

\subsection{Experimental Environment}
We construct our neural networks within the TensorFlow (2.8.1) framework\footnote{https://www.tensorflow.org} inside of Python3.
Matlab 2022a is used for PCG.
The system used is an Intel Xeon Silver 4210R that contains 10 physical cores and supports 20 threads.
The CPUs run at 2.4 GHz.
The system contains a total of 64 GB of DDR4 (4x16GB 2933MHz).
Training is done with the system's Nvidia Quadro RTX4000 GPU with 8GB of GDDR6.

\subsection{Network Training}
All networks are trained with a  custom made version of AdaGrad~\cite{adagrad} built with the Tensorflow framework.
The custom made version allows utilizing only the parameters associated with the nonzero structure of $L$.
Moreover, we could optimize the performance of this over the built-in AdaGrad of Tensorflow/Keras\footnote{https://www.tensorflow.org/guide/keras} by utilizing SpMV operations.
A total of $\sqrt{N}$ training vectors were generated, and the models were trained iteratively utilizing a batch of 1 vector per iteration.
This value is based on the theoretical number of iterations for the convergence of iterative solvers.
Stochastic Gradient Descent (SGD) was also considered initially but it required more tuning and iterations to train.
With the sparse optimizations and the reduction of iterations, the AdaGrad choice is cheaper in terms of the training time (though theoretically more expensive in terms of the number of floating-point operations and memory). 
We consider two sets of training.  
In particular, we train with one set of randomly selected $X$ (denoted as NN) and one set with randomly selected $X$ with normalized samples (denoted as NNN).
We set the AdaGrad parameter to be $\alpha = N^{3/2}/20000$ for samples that are normalized, and the AdaGrad parameter to be $\alpha = 0.1$ for samples that are not normalized.
The loss function utilized was the mean square error (MSE) as shown in Equation~\eqref{eq:mse}.

%\begin{figure*}[tbh]
%    \centering
%    \includegraphics[width = 1.0\textwidth]{photo/norcm.png}
%    \caption{Number of iterations to converge with utilizing natural ordering.}
%    \label{fig:norcm}
%\end{figure*}

\begin{figure}[h]
\centering
\subfloat[First Half, NNZ $<$ 5,000k]{\includegraphics[width=.48\textwidth]{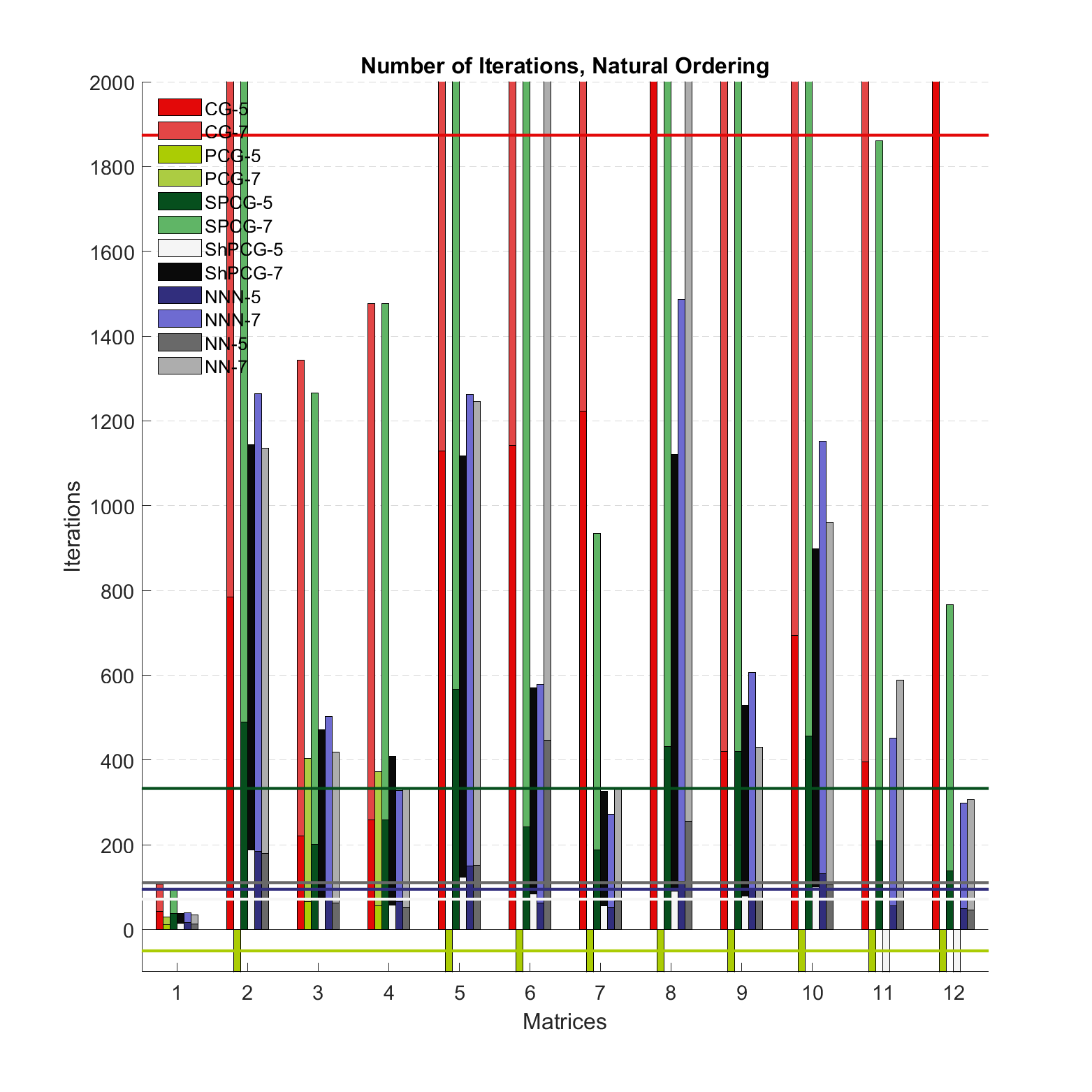}}\\
\subfloat[Second Half, NNZ $>$ 5,000k]{\includegraphics[width=.48\textwidth]{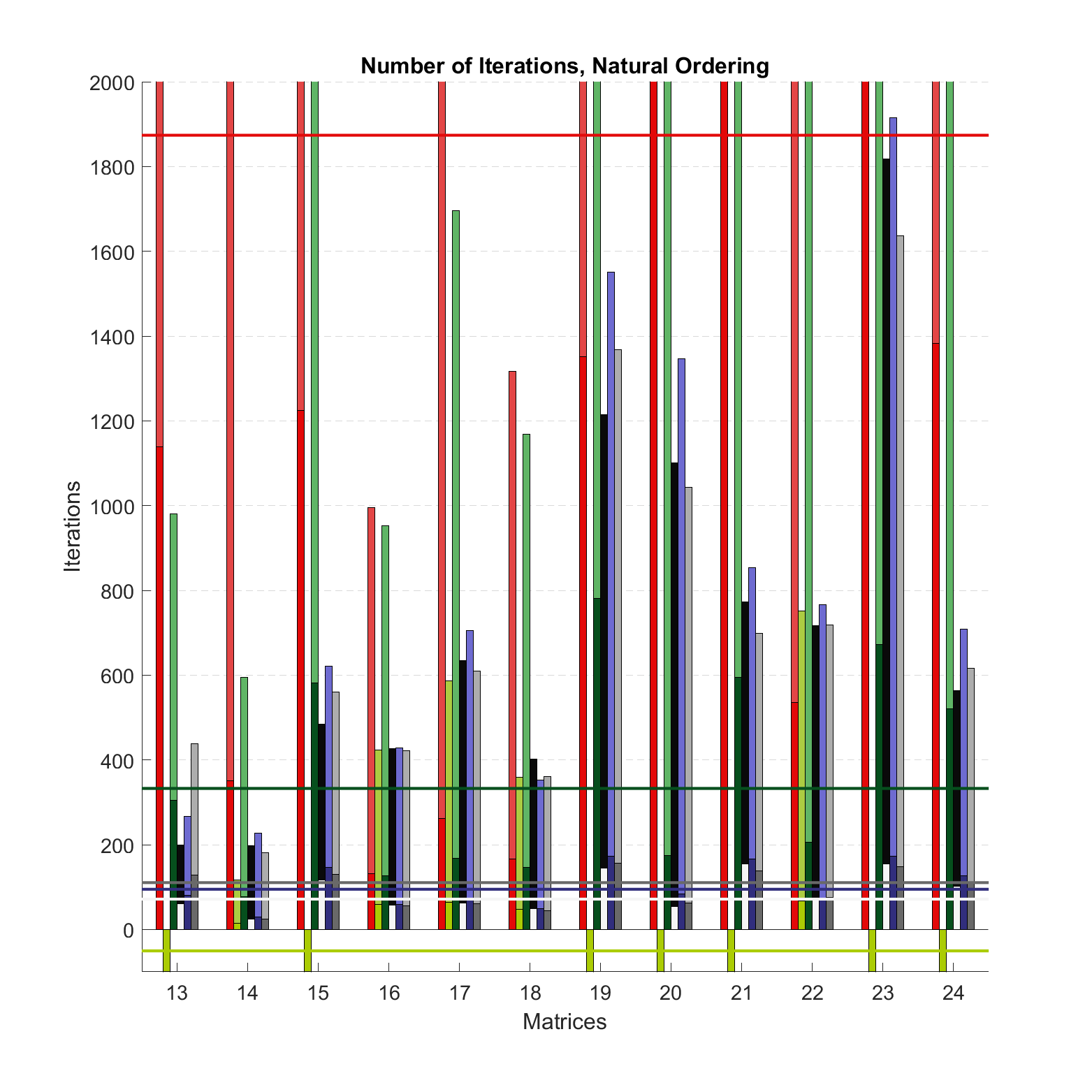}}
\caption{Number of iterations to converge to a solution when the sparse matrix is ordered in their natural ordering. The bars represent the raw number of iterations and the lines represent the average iteration for the method across all 24 matrices. In many cases, tradition \texttt{PCG} fails because the incomplete factorization fails.  In several cases, even scaling with shifting \texttt{ShCG} fails.  The only method that works for all cases while constantly reducing iteration count is the two neural network based methods.}
\label{fig:norcm}
\end{figure}

\section{Experimental Evaluation of Quality}
\label{sec:exp}
In this section, we evaluate the quality of the incomplete Cholesky factorization generated using the neural network model.
To evaluate the quality, we consider the number of iterations for PCG to converge.
While the timing performance of PCG is important, this is primarily dominated by the number of iterations, and therefore the number of iterations is a sufficient measure even in parallel execution.

We consider the quality of two different neural network models.
The first is trained with normalized sample vectors and an AdaGrad parameter of $\alpha = N^{3/2}/20000$.
We denote this one as \nnn.
The second is trained using non-normalized sample vectors and an AdaGrad parameter of $\alpha = 0.1$.
We denote this one as \nn.
Data normalization is very common in training large neural networks. For our model, we desire to test its impact, and the significance is twofold.
The first is that normalized sample vectors would provide a guaranteed larger space that is spanned by the $\sqrt{N}$ sample vectors and it may impact the training time as we are considering the transformation onto this vector in each training iteration.
The second is that normalized sample vectors would provide a smaller distribution of values that are more ideal to the numerical properties of optimization methods used to train neural networks like AdaGrad.
However, this normalization takes some small amount of time and may not capture some larger effect (e.g., extreme scaling found in matrices from multi-physics problems).

We compare our two models against those in shown Table~\ref{tab:methods}.
All PCG methods utilize a $ICHOL(0)$, i.e., all preconditioners have the same nonzero pattern as the input matrix.
This is also why the number of iterations is a sufficient measure of quality, as the same amount of floating-point operations is done in each iteration for all methods except for \cg,  which does not utilize a preconditioner. 
Our scaled methods (\texttt{SCG} and \texttt{ShCG}) utilize a scaling using the diagonal entries.
We used a value of 0.2 for a diagonal shift after exploring multiple, and found that this works the best on average for the whole test suite. 
We also consider all these methods for two values of relative tolerance (Tol) for convergence, i.e., 1$e$-5 and 1$e$-7.
We tested up to a maximum of 10,000 iterations, though we cut off our figures at 2,000 due to space.
Lastly, we also consider using the natural and RCM orderings.

\begin{table}[tbh]
\centering
    \caption{Methods tested for quality.  All but \texttt{CG} have the same number of floating point operations per iteration.}
    \label{tab:methods}
    \begin{tabular}{|c|c|c|}
    \hline
    \textbf{Name} & \textbf{Description} & \textbf{Tol} \\
    \hline
    \cga &  CG  &  1$e$-5 \\
    \cgb &  CG  & 1$e$-7  \\
    \pcga & PCG with $ICHOL(0)$  & 1$e$-5  \\
    \pcgb & PCG with $ICHOL(0)$  & 1$e$-7 \\
    \scga & PCG with scaled matrix  & 1$e$-5 \\
    \scgb & PCG with scaled matrix  & 1$e$-7 \\
    \shcga & PCG with scaled and shifted (.2) & 1$e$-5 \\
    \shcgb & PCG with scaled and shifted (.2) & 1$e$-7 \\
    \nnna & PCG with NN generated $L$ and normalized samples & 1$e$-5\\
    \nnnb & PCG with NN generated $L$ and normalized samples & 1$e$-7\\
    \nna & PCG with NN generated $L$  & 1$e$-5 \\
    \nnb & PCG with NN generated $L$ & 1$e$-7 \\
    \hline
    \end{tabular}
\end{table}

Figure~\ref{fig:norcm} provides the iteration count for each of the methods tested grouped by the sparse matrices.
The methods with different relative tolerances for convergence are stacked. 
In cases where the method would not converge, the number of iterations is reported as $-100$.
The figure also provides lines plotting the average number of iterations for all sparse matrices with the relative tolerance of 1$e$-5.
In particular, these values are: \cga $\sim$ 1874; \scga $\sim$ 333; \shcga $\sim$ 72; \nnna $\sim$ 95; \nna $\sim$ 110.
We mark \pcg below the line as it fails to converge for almost all sparse matrices, and the average of converging cases provides an inaccurate visual account of its performance. 
We first notice that our neural network models (i.e., $NN$ and $NNN$) are the only preconditioning method besides \scg that converges for all sparse matrices in the test suite.
As mentioned in the background (Section \ref{sec:background}), the development of a preconditioner is as much an art as a science that is guided by expert experience and trials (e.g., \shcg failing for sparse matrices 11 and 12 despite being the best method for many of the other sparse matrices). 
This demonstrates that the use of neural acceleration can help convert this artful practice into a standard function call.

Not only does our neural acceleration method convert it to a simple standard function call that will converge, but it also provides a high-quality preconditioner.
The only method that does better in terms of average iteration count than the two neural acceleration generated preconditioners is the scaled shifted method (\shcg). 
However, this method has its own issues that include not converging for two sparse matrices and having to find an $\alpha$ value that works~\cite{shiftcg}. 
For a couple of matrices (i.e., 4 and 9), the neural network model does better than \shcg.
However, there are a couple of cases where the neural network models can be worse.
These cases include matrices 6 (\texttt{vanbody}) and 20 (\texttt{StocF-1465}).
In the case of matrix 6 (\texttt{vanbody}), a model with the non-normalized sample vectors does not do well, i.e., $\nnb > 4,000$ compared to \cgb $>10,000$ and $\shcgb < 600$  when the tolerance is 1e-7.
However, the model with normalized sample vectors is about on par with \shcg.
This might indicate a scaling issue with the optimization method for training.
On the other hand, for matrix 20 (\texttt{StocF-1465}), one neural network model (\nnn) does poorly and one does well (\nn).

When comparing the two neural network methods (i.e., the one trained with normalized samples (\nnn) and the one trained without normalized samples (\nn)), there is little difference on average.
However, if you look at the performance case-by-case, you can exclude matrices 6 (\texttt{vanbody}) and 13 (\texttt{consph}) and notice that \nn generally does better than \nnn.
This is a nice finding for two particular reasons.
First, normalization does not need to be done, thus saving time in training.
Second, the training of \nn is much less sensitive. 
In particular, a simple parameter of $\alpha = 0.1$ works well with AdaGrad for training, while the training of \nnn is much more difficult.
In the next section, we will compare the time to train the models against the standard methods.

%\begin{figure*}[tbh]
%    \centering
%    \includegraphics[width = 1.0\textwidth]{photo/rcm.png}
%    \caption{The number of iterations needed to converge when input is reordered with RCM.}
%    \label{fig:rcm}
%\end{figure*}

\begin{figure}[h]
\centering
\subfloat[First Half, NNZ $<$ 5,000k]{\includegraphics[width=.48\textwidth]{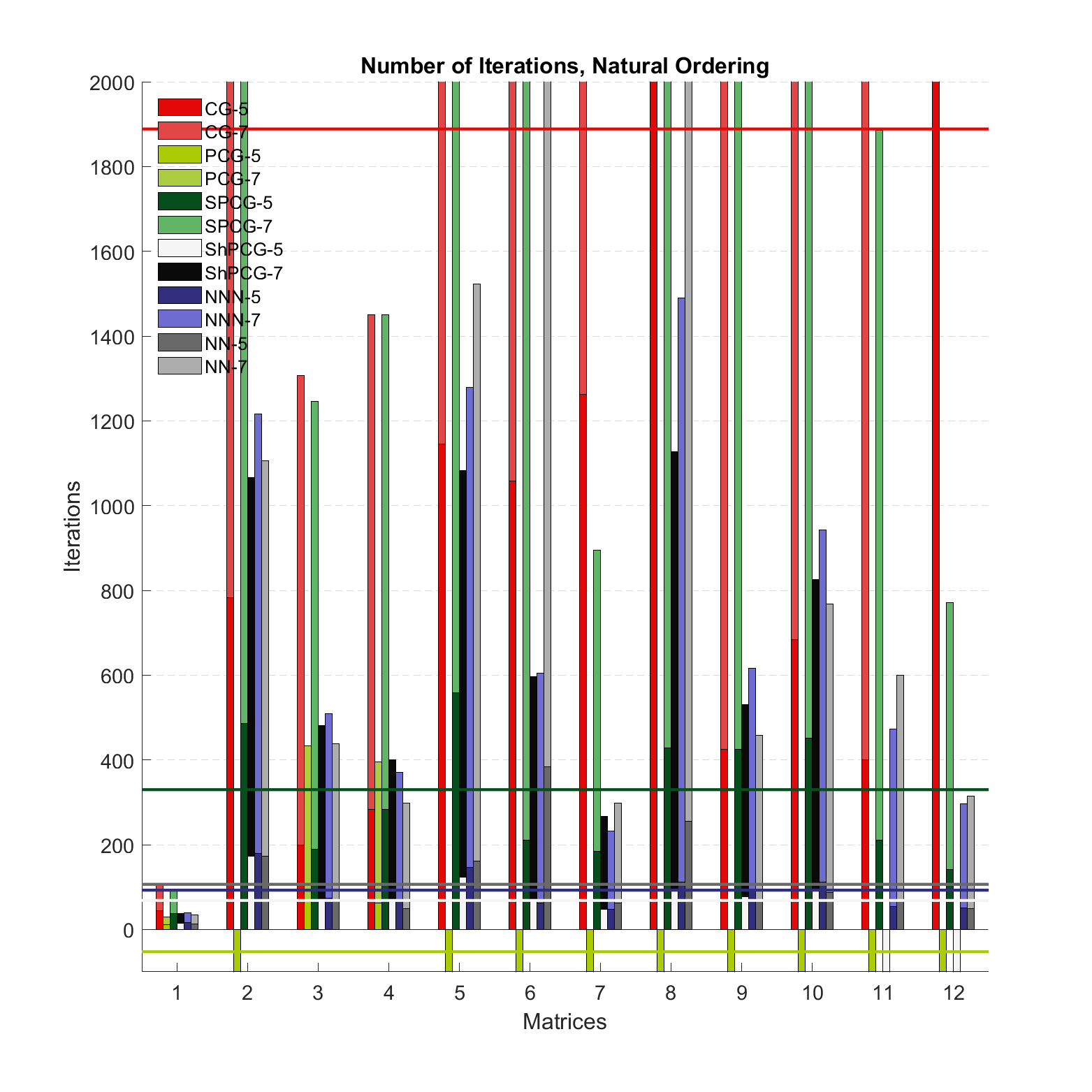}}\\
\subfloat[Second Half, NNZ $>$ 5,000k]{\includegraphics[width=.48\textwidth]{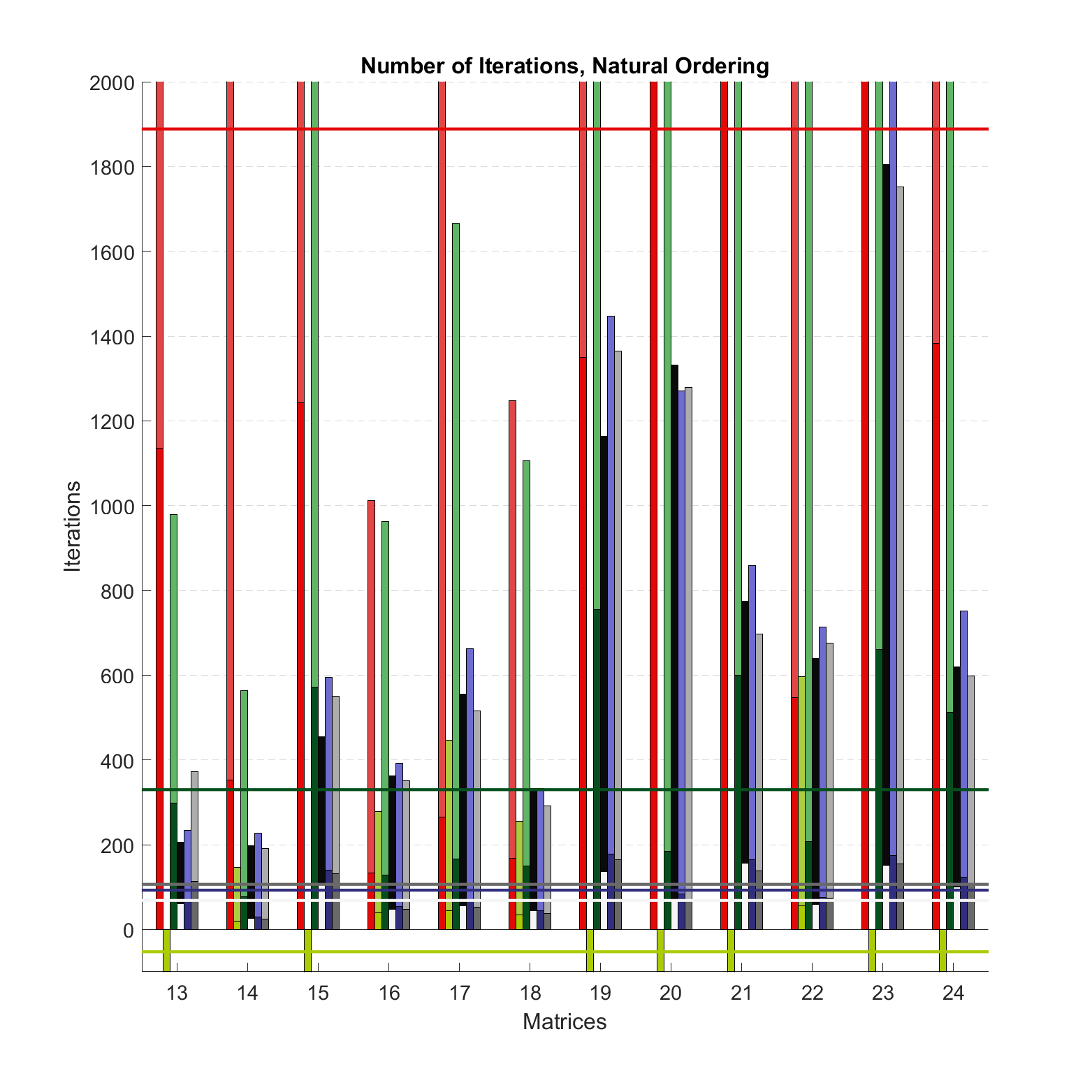}}
\caption{Number of iterations to converge to a solution when the sparse matrix is ordered in the RCM ordering. The bars represent the raw number of iterations and the lines represent the average iteration for the method across all 24 matrices. We notice that the ordering does not seem to impact the number of iterations required by our method. }
\label{fig:rcm}
\end{figure}

For completeness, we also consider the case when the sparse matrices are reordered with RCM.
Figure~\ref{fig:rcm} presents the number of iterations for sparse matrices ordered with RCM.
Overall, the iteration counts for the natural and RCM orderings are about the same, even though we know in theory this may not always be true~\cite{duff}.
With RCM ordering, the averages are: \cga $\sim$ 1889; \scga $\sim$ 330; \shcga $\sim$ 69; \nnna $\sim$ 93; \nna $\sim$ 106.
We do note that the time per iteration may be less for sparse matrices ordered with RCM because of their spacial locality in memory~\cite{csrk,csrkgpu}.
However, this does demonstrate that, for our test suite, the quality of the neural network models seems to depend on ordering.
In future work, we will dive deeper into this because of the importance it might have in GPU computations.
In particular, many GPUs require formats and orderings that are different from those on multicore CPUs to achieve high performance~\cite{csrkgpu}.
If ordering does not matter with these preconditioners, they may be better on GPUs than methods that are dependent on ordering.

\begin{figure}[tbh]
\centering
\includegraphics[width=1\linewidth]{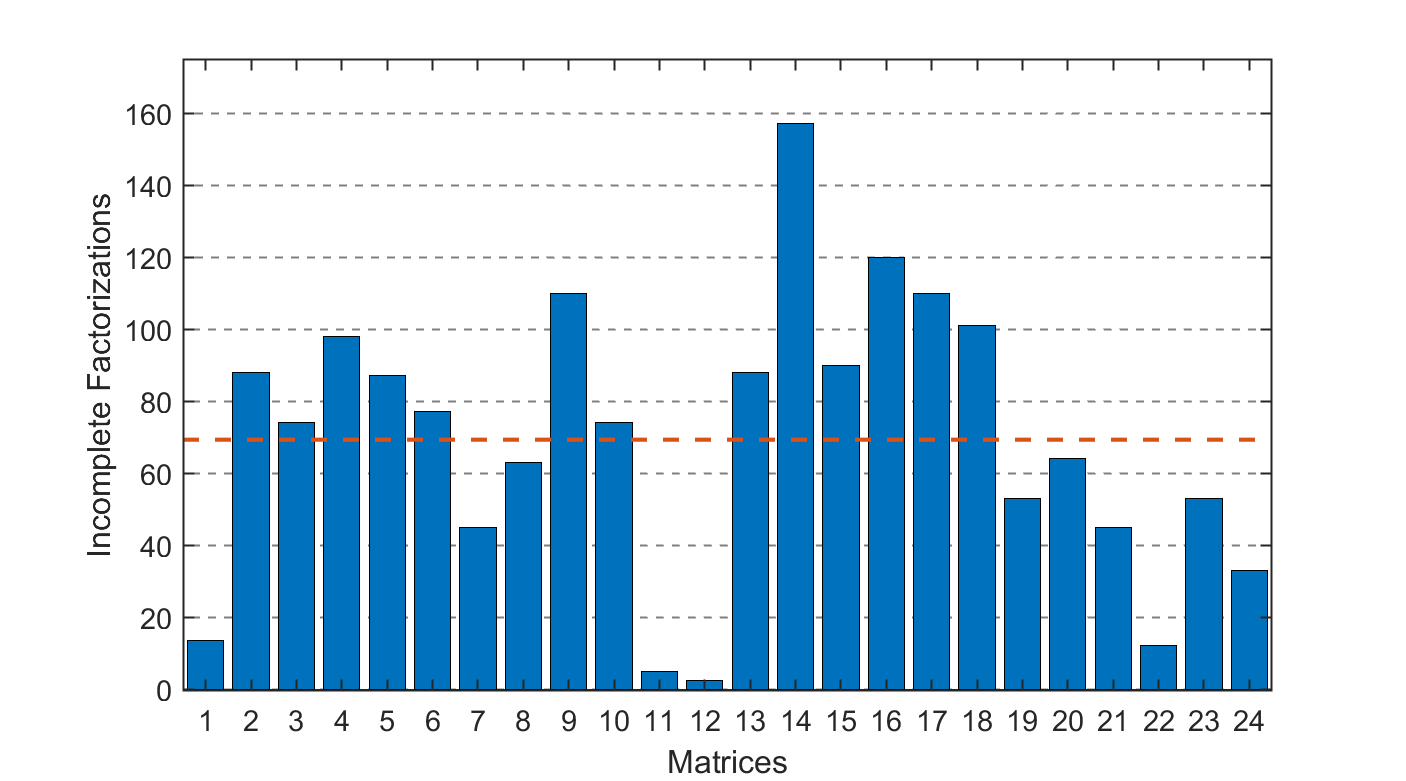}
\caption{Evaluation of the cost in terms of the number of factorizations ($NFacts(M)$). Each bar represents the value of $NFacts$ for the particular sparse matrix and the dotted line represents the average. }
\label{fig:time}
\end{figure}

\section{Experimental Evaluation of Cost}
\label{sec:cost}
Lastly, we evaluate the cost of generating the preconditioners using the neural acceleration method in comparison with the traditional methods (e.g., utilizing $ICHOL$ without scaling or just scaling).
The question is about what is the timing cost of training such a preconditioner.
Normally the cost of utilizing a neural network is dominated by two factors: (a) time for training; and (b) time for searching for the correct set of hyperparameters utilized in backpropagation and regularization.
Since our neural network model is so simple, a search of the hyperparameter space is not needed, and only training matters.
However, this type of training can be very expensive as it requires numerical optimization.
We note that many of the current GPU accelerators are optimized for this type of calculation with new progress coming for sparse applications due to graph neural networks.
To gauge the cost we use the following metric:
\begin{equation}
NFacts(M) = \frac{TimeNN(M)}{TimeIChol(M)}.
\end{equation}

This metric aims to gauge the number of incomplete factorizations for a sparse matrix ($M$) that can be completed in the time for training the neural network ($TimeNN(M)$) if the time per incomplete factorization is $TimeIChol(M)$.
The value of $TimeIChol(M)$ is taken from a modified version of Javelin~\cite{javelin} for the fastest case that will factor without losing numerical stability.  
The Javelin package is a highly efficient package for incomplete factorization that utilizes threads in a shared memory environment.
We have modified Javelin to do $ICHOL(0)$ in place of being designed for incomplete LU.
Moreover, we report the time for using 4 threads with Javelin as this was the largest number of threads that did not suffer from Amdahl's law for all matrices in the test suite.
We do not consider the time for moving data in both $TimeNN(M)$ and $TimeIChol(M)$.
We justify the use of this metric as follows.
The neural acceleration method using our neural networks would only train one model, while someone trying to find a ``good" preconditioner utilizing different methods of scaling and parameters for shifting would try multiple preconditioners to match their input case.
The metric judges how many of these cases could the user try in the time that our neural acceleration method trains the preconditioner, without user input.
Note that there are a couple of factors that may make this metric less than realistic.
The first is that a failed incomplete factorization may take less time.
The second is that a successful incomplete factorization still might not be as optimal.  The only way to test is to apply PCG and observe the number of iterations.

Figure~\ref{fig:time} presents the results of using this metric for our test suite.
We note that on average the cost is about 69.3 incomplete factorizations.
In some cases, e.g., matrix 14 (\texttt{G3\_circuit}), this value can be much higher.
On the other hand, several matrices have much smaller cost, e.g., matrix 1 (\texttt{minsufo}), matrix 11 (\texttt{cant}), matrix 12 (\texttt{shipsec5}), and matrix 22 (\texttt{inline\_1}).
One factor that seems to dominate this trade-off is the density of the matrix.
Since we utilized a less-than-optimal optimizer (i.e., our version of AdaGrad), a great deal more work is being done for the more sparse cases with the optimizer than for the sparse incomplete factorization. 
Therefore, the neural acceleration method may consider the density of the input and available hardware (e.g., does the current GPU support sparse tensor operation well?) to determine if a neural network model or the original function call should be used. 

Despite many cases where a traditional search method with $ICHOL$ could be faster, our method is successful as a neural acceleration method overall.
The reason is that a high-quality approximation is found with little to no user interaction in a time that would fit into the compile time of a large scientific application. 
We note that a large scientific application could take more than 5 minutes to compile considering large frameworks like Trilinos\footnote{https://trilinos.github.io/}.
We find that the maximum time to generate a preconditioner using our method is relatively small (i.e., $\sim 117$ second) and the average time is less than a minute (i.e., $\sim 53$ seconds) using our GPU.

%\section{Future Work}
%\label{sec:future}
%We plan to extend our work in this area by considering several directions.
%The first is a stronger comparison of quality and time trade-off with other parameters.
%Though our method works well as a neural acceleration method, finding better parameters could aid in our development of a faster library call.
%In particular, we want to expand our test suite and consider more GPU-friendly orderings. 
%The second is the extension and testing of this method for incomplete $LU$.
%This extension has some considerations that need to be addressed as we outlined in Section~\ref{sec:discussion}.
%However, this may be even more impactful as the issue of designing good preconditioners for GMRES can be even more sensitive. 
%Lastly, we will extend our comparison to the theoretical preconditioners that utilize sparsification~\cite{grass}.
%Although these tend to not work well in practice, insights might be gained by directly comparing the preconditioners generated by the theoretical method and our neural network method.

\section{Conclusion}
\label{sec:con}
In this work, we developed a neural network modeling method for incomplete Cholesky factorization that can be utilized for neural acceleration.
The goal of this method is to produce a good and inexpensive approximation that could be computed at compile time or in parallel on a GPU during execution. 
The incomplete Cholesky factorization method is an ideal method for neural acceleration to approximate the input matrix $A$ using the iterative method of PCG.
In doing so, we develop a simple two-layer sparse artificial neural network model that utilizes a straightforward implementation of AdaGrad to train.
No meta parameters related to regularization of dropout are needed.
As a result, the model is as simple and cheap as expected. 
In particular, we demonstrated that a model as good as a standard model could be computed with only $\sqrt{N}$ samples and iterations (i.e., the expected number for this type of training).
Not only was the method efficient in training, but it was also the only method that is able to provide a consistent decrease in iteration count for the whole test suite.
As such, the method works as a black box preconditioner that would be ideal in cases where the application user does not have insight into the problem. 

\section*{Acknowledgment}
This work was supported by NSF-2044633, CAREER: Fast, Energy Efficient Irregular Kernels via Neural Acceleration.\\
We would like to thank Haun Le who worked as an undergraduate researcher developing some of the initial programming framework.

\ifCLASSOPTIONcaptionsoff
  \newpage
\fi

\bibliographystyle{IEEEtran}
\bibliography{Submit}

\vfill


\end{document}